\documentclass[11pt,a4paper]{article}
\usepackage[left=15mm, top=10mm, right=15mm, bottom=20mm, nohead=true, nofoot=true]{geometry}

\usepackage[utf8]{inputenc}
\usepackage{authblk}
\usepackage{indentfirst}
\usepackage{misccorr}
\usepackage{graphicx}
\usepackage{amsmath}
\usepackage{amssymb}
\usepackage{multicol}
\usepackage{bm}
\usepackage{longtable}
\usepackage{pdflscape}
\usepackage{epstopdf}
\usepackage{multirow}
\usepackage{adjustbox}
\usepackage{amsfonts}
\usepackage{ifthen}
\usepackage{fancyhdr}
\usepackage{setspace}
\usepackage{xcolor}
\usepackage[round,authoryear,comma,sort&compress,numbers]{natbib}
\usepackage[toc,title,page]{appendix}
\usepackage[hidelinks]{hyperref}
\usepackage{url}

\hypersetup{
    colorlinks=true,
    linkcolor=green,
    citecolor=blue,
    filecolor=magenta,
    urlcolor=cyan,
    pdftitle={Sharelatex Example},
    pdfpagemode=FullScreen,
}

\fancypagestyle{plain}{\chead{\texttt{\textcolor{brown}{Communications of BAO, Vol. 71, Issue 2, 2024, pp. 394-397}}}}
\title{\textbf{Star Formation History of the Local Group Dwarf Irregular Galaxy, NGC\,6822}}
\author[1]{Khatamsaz F. \thanks{fate.khatamsaz@gmail.com, Corresponding author}}
\author[1]{Abdollahi M. \thanks{m.abdollahi@ipm.ir}}
\author[1,2]{Abdollahi H. \thanks{Hediehabdollahi@ipm.ir}}
\author[1]{Javadi A.\thanks{atefeh@ipm.ir}}
\author[3]{van Loon J. Th. \thanks{j.t.van.loon@keele.ac.uk}}
\affil[1]{\scriptsize School of Astronomy, Institute for Research in Fundamental Sciences (IPM), Tehran, 19568-36613, Iran}
\affil[2]{Konkoly Observatory, HUN-REN Research Centre for Astronomy and Earth Sciences, MTA Centre of Excellence, Konkoly-Thege Mikl\'os \'ut 15-17, H-1121, Budapest, Hungary}
\affil[3]{\scriptsize Astrophysics Group, Lennard-Jones Laboratories, Keele University, Staffordshire ST5 5BG, United Kingdom}

\begin{document}
\pagestyle{empty}
\newpage
\pagestyle{fancy}
\label{firstpage}
\date{}
\maketitle

\begin{abstract}
NGC\,6822 is an isolated dwarf irregular galaxy in the local group at a distance of $\sim$ 490 kpc. In this paper, we
present the star formation history (SFH) within a field with a radius of $\sim$ 3 kpc, beyond the optical body of the
galaxy ( $\sim$ 1.2 kpc). We utilized a novel method based on evolved asymptotic giant branch (AGB) stars. We collected the Near-infrared data of 329 variable stars, including long-period and -amplitude variables and Carbon-rich AGB stars. We used stellar evolutionary track and theoretical isochrones to obtain the birth mass, age, and pulsation duration of the detected stars to calculate the star formation rate (SFR) and trace the SFH of the galaxy. We studied the galaxy’s star formation history for the mean metallicity value of Z $\approx$ 0.003. We reconstructed the SFH for two regions: the bar region, a central rectangular area, and the outer region, which covers a circular field beyond the bar region and extends to a radius of 3 kpc. Our results show a significant burst of star formation around 2.6 and 2.9 Gyr ago in the bar and outer regions, respectively. Additionally, we observed a notable enhancement in the SFR in the bar region over the past 200 Myr.

\end{abstract}

\section{Introduction}
Investigating the star formation history (SFH) of galaxies provides profound insights into their formation and evolution.
NGC\,6822 is an isolated dwarf irregular (dIrr) galaxy within the local group, located at a distance of $\sim$ 490 kpc (\citealp{lee1993tip}; \citealp{mateo1998dwarf}) in constellation Sagittarius. The exotic structure of NGC\,6822 features a central bright bar, stretching from North to South (\citealp{hodge1977structure}; \citealp{hodge1991cosmos}). The central bar is embedded in an HI envelop, extending from the northwest (NW) to the southeast (SE); this central structure is surrounded by a extensive elliptical halo located at a radial distance of $\sim$ 12 kpc (\citealp{de2000evidence}; \citealp{hwang2005discovery}; \citealp{zhang2021panoramic}). Despite its gas-richness, NGC\,6822 has a relatively low metallicity. While a wide range of metallicities have been associated to various ages of the galaxy, the studies by \citealp{tolstoy2001using} and \citealp{davidge2003metallicity} reported a mean value of [Fe/H] $\approx - 1.00$ dex (Z $\approx$ 0.003), derived based on the red giant branch (RGB) Ca II triplet and the slope of RGB (J, K bands), respectively. 

The evolved stellar population of NGC\,6822, including asymptotic giant branch (AGB; \citealp{marigo2008evolution}) and red supergiant (RSG) stars, are excellent implements to trace the SFH over periods spanning from a few million years to 10 billion years (\citealp{ekstrom2013red}). The significant brightness of AGBs ($\sim$ 10$^{3}-$10$^{4}$ L$_{\odot}$; \citealp{hofner2018mass}), along with their radial pulsations makes these stars conveniently detectable in infrared wavelengths. In this work, we apply a method developed by Javadi et.\:al.\:(\citeyear{javadi2011auk}, \citeyear{javadi2011uk}, \citeyear{javadi2013uk}, \citeyear{javadi2015uk}) based on AGBs that pulsate with periods longer than a 100 days, known as long-period variable (LPV) stars (\citealp{iben1983asymptotic}; \citealp{whitelock2003obscured}). The SFH of many nearby galaxies in the LG has been derived utilizing this method (\citealp{rezaei2014star}; \citealp{javadi2016role}, \citealp{javadi2017uk}; \citealp{hamedani2017long}; \citealp{hashemi2019evolved}; \citealp{navabi2021isaac}; \citealp{saremi2021isaac}; \citealp{abdollahi2023isaac}; \citealp{aghdam2024complex}); in the following, we will apply this method using the evolved stellar population of NGC\,6822 to reconstruct its SFH.

\vspace{10pt}
\section{Data and Method}

The sample used for derivation of the SFH contains the data of 329 evolved stars in J, H, and K$_s$ bands, selected and combined from several catalogs published by \citealp{kacharov2012spectra},\citealp{whitelock2013local}, and Sibbons\:et.\:al.(\citeyear{sibbons2012agb}, \citeyear{sibbons2015spectral}). Most of the stars in the sample, including 228 long-period and long-amplitude variables and spectroscopy-confirmed Carbon-rich AGB stars, are distributed within the bar region. This region is defined as a rectangular area of 9 $\times$ 21 arcmin$^{2}$, situated in the center of the galaxy. Meanwhile, the remaining 101 stars are only spectroscopy-confirmed Carbon-rich AGBs dispersed beyond the bar region, extending up to a radius of 3 kpc, which we refer to as the outer region throughout this paper. We must note that due to insufficient available data, the outer region stars are solely spectroscopy-confirmed Carbon-rich AGBs. As a result, this limitation restricts and narrows the age range of the derived SFH associated with this area.

In order to obtain the SFH, we calculated the star formation rate (SFR), $\xi(t)$ (M$_\odot$\,yr$^{-1}$), which is defined as the mass of gas that has converted into stellar mass over a specific time interval. To do so, we used the Padova stellar evolutionary tracks and isochrones (\citealp{marigo2008evolution}, \citeyear{marigo2017new}), assuming constant metallicities, to relate the magnitude of each star to its birth mass. Similarly, we obtained the age and pulsation duration (when LPVs are in pulsating phase). Subsequently, we sorted the stars based on age and divided them into several bins. Then, we derived the SFR for each bin, with its associated age and mass range, using the following relation:

\begin{equation}
\label{eq:1}
  \xi(t) = \frac{\int_{\mathrm{min}}^{\mathrm{max}} f_{\mathrm{\mathrm{IMF}}}(m)\, m\, \mathrm{d}m}%
           {\int_{m(t)}^{m(t+\mathrm{d}t)} f_{\mathrm{IMF}}(m)\, \mathrm{d}m}  \: \frac{\mathrm{d}n^\prime(t)}{\delta (t)},
\end{equation}

where the m is mass, $\mathrm{d}n^{\prime}$ is the number of observed LPVs in each bins, $\delta (t)$ is the pulsation duration, and $f_{\mathrm{IMF}}$ is the initial mass function (IMF) (\citealp{kroupa2001variation}). We also consider a statistical error for each bin, derived based on the Poisson statistics:

\begin{equation}
  \sigma_{\xi(t)} = \frac{\sqrt{N}}%
           {N} \: \xi(t),
\end{equation}

where $N$ is the number of LPVs in each bin.

\section{Results}

We calculated the SFRs in two regions of NGC\,6822: the bar region and the outer region, using the method and the dataset explained in Section\:2. Additionally, we assumed that the metallicity remained constant over time and derived the SFH adopting the mean metallicity of [Fe/H] $=$ $-$1.05 dex (Z $\approx$ 0.003). The utilized model to obtain the parameters required to calculate the SFR were obtained from \citealp{khatamsaz2024investigating} and {\color{blue}{Khatamsaz et al., in preparation}}.

\begin{figure*}[t!] 
    \centering 
    \includegraphics[width=80mm] {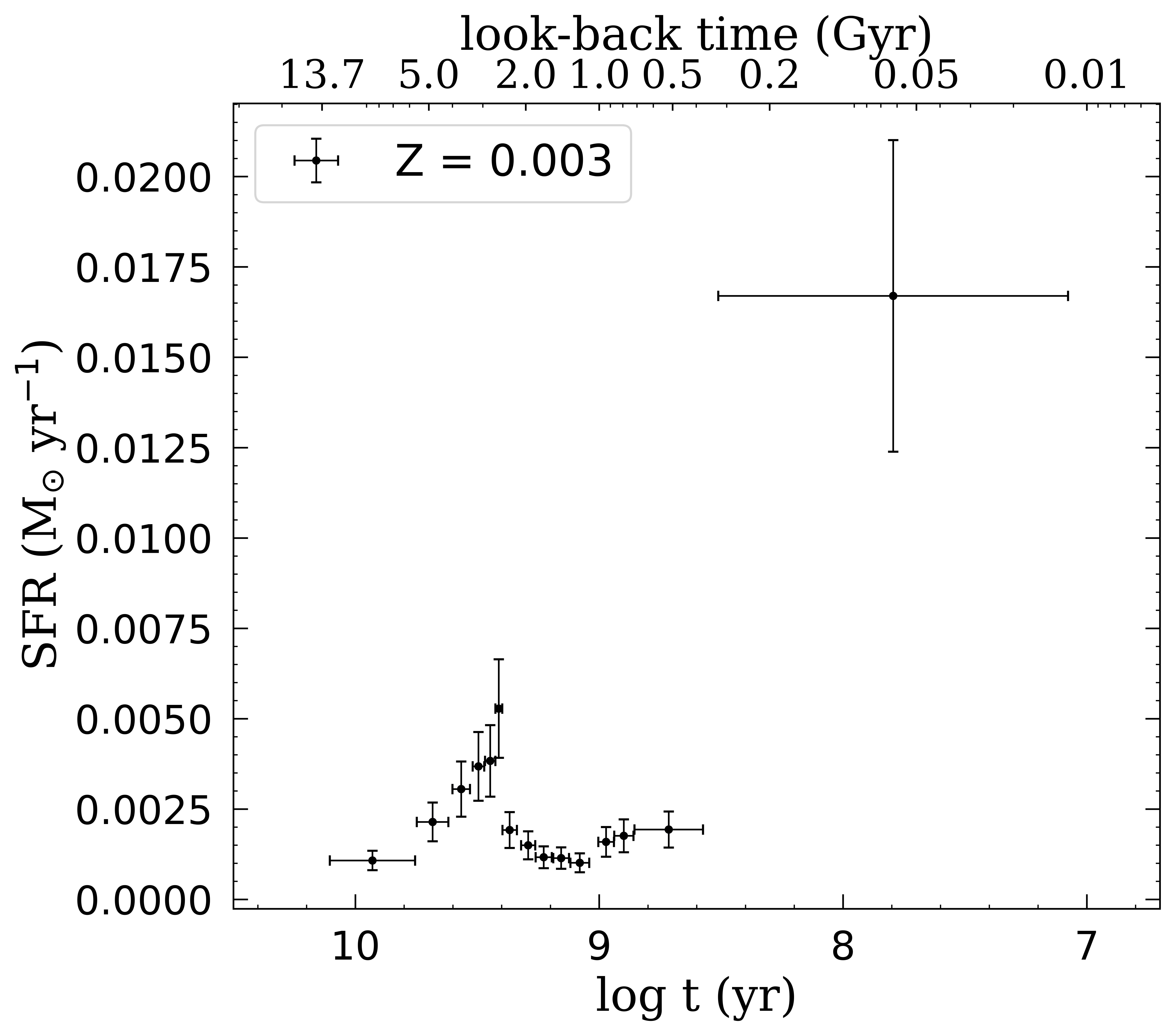} 
     \hspace{5pt}
    \includegraphics[width=80mm] {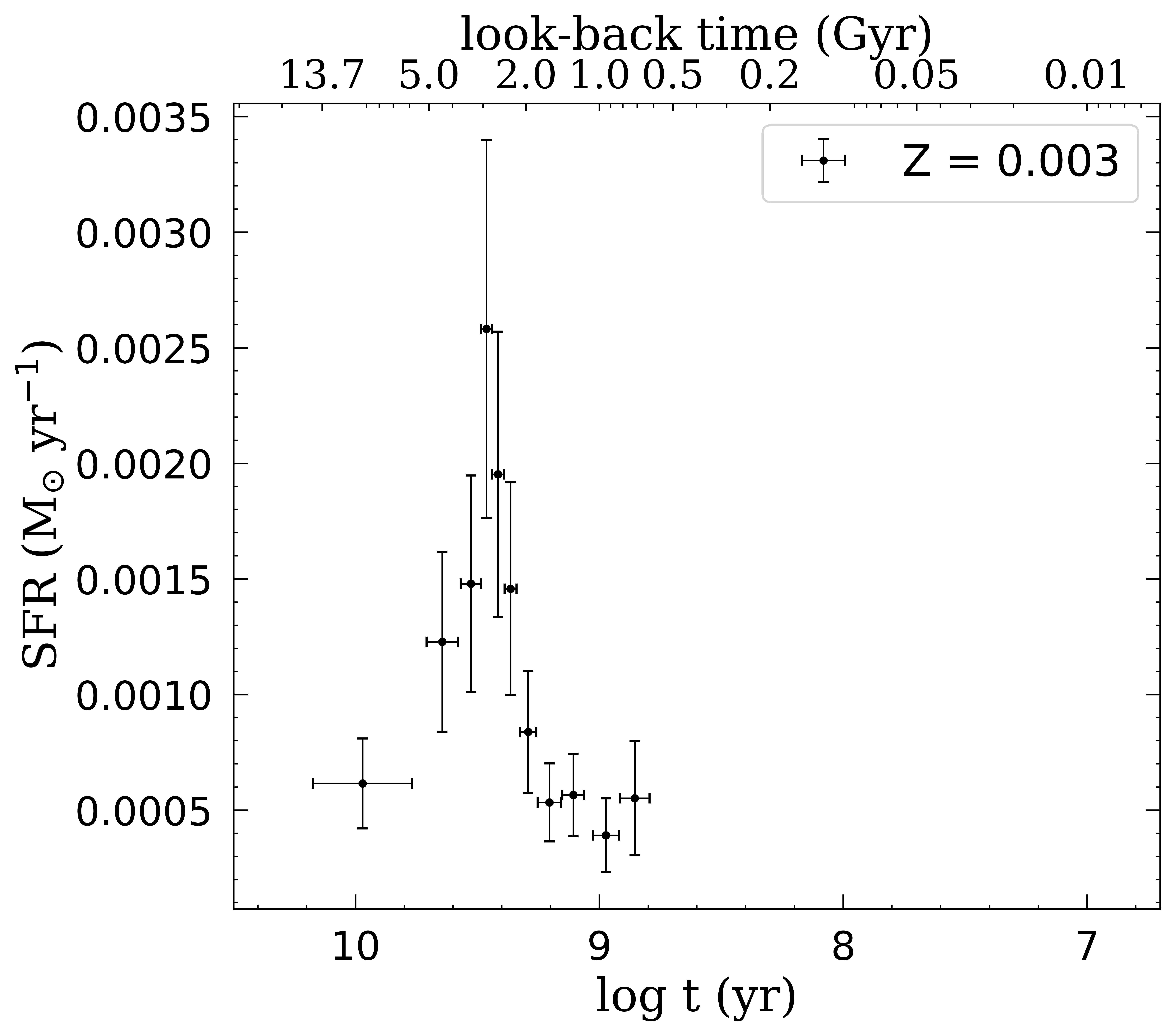} 
    \caption{The SFH in NGC\,6822 with the adoption of the mean metallicity of Z $\approx$ 0.003. The left panel presents the results for the bar region, and the results for the outer region are displayed on the right panel.} \label{fig:Fig.1} 
\end{figure*}

The left panel of Fig.\,\ref{fig:Fig.1} shows the results for the bar region. In this area, the star formation begins as early as $\sim$ 12.7 Gyr ago, with a rate of 1.1 $\pm$ 0.3 $\times$ $10^{-3}$ M$_{\odot}$\,yr$^{-1}$. Following this, the SFR increases gradually for $\sim$ 10.0 Gyr and peaks at $\sim$ 2.6 Gyr ago, reaching a rate of 5.3 $\pm$ 1.4 $\times$ $10^{-3}$ M$_{\odot}$\,yr$^{-1}$. The star formation in the bar region then decreases for $\sim$ 1.6 Gyr. However, it begins to increase once again in the past $\sim$ 1.0 Gyr and experiences a significant enhancement in its rate, reaching the maximum rate of $\sim$ 17.0 $\pm$ 4.3 $\times$ $10^{-3}$ M$_{\odot}$\,yr$^{-1}$ over the past 300 Myr. This rate is in good agreement with the recent SFR of 21.0 $\times$ $10^{-3}$ M$_\odot$\,yr$^{-1}$ derived by \citealp{Hodge1993eso} based on the H$\alpha$ luminosity.

As we mentioned previously, the dataset of the outer region, as defined in Section\,2, is limited to Carbon-rich AGB stars. Therefore, our results are confined to the age range of the used sample, which falls within the range of 15.0 Gyr $<$ look-back time $<$ 620 Myr, considering the mean metallicity of Z $\approx$ 0.003. The right panel in Fig\,\ref{fig:Fig.1} presents the results for the outer region. Similar to the bar region, the SFR peaks at $\sim$ 2.9 Gyr ago, reaching the maximum rate of $\sim$ 2.6 $\pm$ 0.8 $\times$ $10^{-3}$ M$_{\odot}$\,yr$^{-1}$, which is roughly as half as the rate obtained for the same epoch of star formation burst in the bar region. Following this, the SFR gradually decreases up until $\sim$ 830 Myr ago. However, as mentioned before, due to the limitation of data in the outer region, we cannot retrieve any results for ages younger than $\sim$ 620 Myr ago. Consequently, it remains unclear whether the most recent bin indicates the initiation of a new epoch of star formation similar to the one observed in the bar region.

\section{Conclusion}

In this paper, we applied a novel method to evolved AGB
stars to find the SFH of NGC\,6822. Our results show that the
SFR in the galaxy has increased significantly
during the last 200 Myr in the bar region. Additionally, the star formation bursts and peaks $\sim$ 2.6 and $\sim$ 2.9 Gyr in the bar and outer regions, respectively. The presence of this peak provides evidence for the tidal interaction of NGC\,6822 and the Milky Way proposed by \citealp{zhang2021panoramic}. Furthermore, the non-uniform SFH shows that despite the noticeable isolation of the galaxy, it has gone under events that triggered the star formation activities. Our upcoming paper on the SFH of NGC\,6822 will discuss the subject further and present a new plausible scenario to explain the recent unusual burst of star formation in this galaxy.

\scriptsize
\bibliographystyle{ComBAO}
\nocite{*}
\bibliography{references}

\begin{thebibliography}{}
\makeatletter
\relax
\def\mn@urlcharsother{\let\do\@makeother \do\$\do\&\do\#\do\^\do\_\do\%\do\~}
\def\mn@doi{\begingroup\mn@urlcharsother \@ifnextchar [ {\mn@doi@} {\mn@doi@[]}}
\def\mn@doi@[#1]#2{\def\@tempa{#1}\ifx\@tempa\@empty \href {http://dx.doi.org/#2} {doi:#2}\else \href {http://dx.doi.org/#2} {#1}\fi \endgroup}
\def\mn@eprint#1#2{\mn@eprint@#1:#2::\@nil}
\def\mn@eprint@arXiv#1{\href {http://arxiv.org/abs/#1} {{\tt arXiv:#1}}}
\def\mn@eprint@dblp#1{\href {http://dblp.uni-trier.de/rec/bibtex/#1.xml} {dblp:#1}}
\def\mn@eprint@#1:#2:#3:#4\@nil{\def\@tempa {#1}\def\@tempb {#2}\def\@tempc {#3}\ifx \@tempc \@empty \let \@tempc \@tempb \let \@tempb \@tempa \fi \ifx \@tempb \@empty \def\@tempb {arXiv}\fi \@ifundefined {mn@eprint@\@tempb}{\@tempb:\@tempc}{\expandafter \expandafter \csname mn@eprint@\@tempb\endcsname \expandafter{\@tempc}}}

\bibitem[\protect\citeauthoryear{Abdollahi et~al.,}{Abdollahi et~al.}{2023}]{abdollahi2023isaac}
Abdollahi H.,  et~al., 2023, The Astrophysical Journal, 948, 63

\bibitem[\protect\citeauthoryear{Aghdam et~al.,}{Aghdam et~al.}{2024}]{aghdam2024complex}
Aghdam S.~T.,  et~al., 2024, The Astrophysical Journal, 972, 47

\bibitem[\protect\citeauthoryear{Davidge}{Davidge}{2003}]{davidge2003metallicity}
Davidge T.,  2003, Publications of the Astronomical Society of the Pacific, 115, 635

\bibitem[\protect\citeauthoryear{De~Blok \& Walter}{De~Blok \& Walter}{2000}]{de2000evidence}
De~Blok W.,  Walter F.,  2000, The Astrophysical Journal, 537, L95

\bibitem[\protect\citeauthoryear{Efremova et~al.,}{Efremova et~al.}{2011}]{efremova2011recent}
Efremova B.~V.,  et~al., 2011, The Astrophysical Journal, 730, 88

\bibitem[\protect\citeauthoryear{Ekstr{\"o}m, Georgy, Meynet, Groh  \& Granada}{Ekstr{\"o}m et~al.}{2013}]{ekstrom2013red}
Ekstr{\"o}m S.,  Georgy C.,  Meynet G.,  Groh J.,   Granada A.,  2013, EAS Publications Series, 60, 31

\bibitem[\protect\citeauthoryear{Hamedani~Golshan, Javadi, van Loon, Khosroshahi  \& Saremi}{Hamedani~Golshan et~al.}{2017}]{hamedani2017long}
Hamedani~Golshan R.,  Javadi A.,  van Loon J.~T.,  Khosroshahi H.,   Saremi E.,  2017, Monthly Notices of the Royal Astronomical Society, 466, 1764

\bibitem[\protect\citeauthoryear{Hashemi, Javadi  \& van Loon}{Hashemi et~al.}{2019}]{hashemi2019evolved}
Hashemi S.~A.,  Javadi A.,   van Loon J.~T.,  2019, Monthly Notices of the Royal Astronomical Society, 483, 4751

\bibitem[\protect\citeauthoryear{Hodge}{Hodge}{1977}]{hodge1977structure}
Hodge P.~W.,  1977, Astrophysical Journal, Suppl. Ser., Vol. 33, p. 69-82, plates 13-16, 33, 69

\bibitem[\protect\citeauthoryear{Hodge}{Hodge}{1993}]{Hodge1993eso}
Hodge P.~W.,  1993, ESO Conference and Workshop Proceedings, No. 49, p. 501

\bibitem[\protect\citeauthoryear{Hodge, Smith, Eskridge, MacGillivray  \& Beard}{Hodge et~al.}{1991}]{hodge1991cosmos}
Hodge P.,  Smith T.,  Eskridge P.,  MacGillivray H.,   Beard S.,  1991, Astrophysical Journal, Part 1 (ISSN 0004-637X), vol. 379, Oct. 1, 1991, p. 621-630., 379, 621

\bibitem[\protect\citeauthoryear{H{\"o}fner \& Olofsson}{H{\"o}fner \& Olofsson}{2018}]{hofner2018mass}
H{\"o}fner S.,  Olofsson H.,  2018, The Astronomy and Astrophysics Review, 26, 1

\bibitem[\protect\citeauthoryear{Hwang et~al.,}{Hwang et~al.}{2005}]{hwang2005discovery}
Hwang N.,  et~al., 2005, Proceedings of the International Astronomical Union, 1, 257

\bibitem[\protect\citeauthoryear{Hwang, Lee, Lee, Park, Park, Kim  \& Park}{Hwang et~al.}{2011}]{hwang2011extended}
Hwang N.,  Lee M.~G.,  Lee J.~C.,  Park W.-K.,  Park H.~S.,  Kim S.~C.,   Park J.-H.,  2011, The Astrophysical Journal, 738, 58

\bibitem[\protect\citeauthoryear{Iben~Jr \& Renzini}{Iben~Jr \& Renzini}{1983}]{iben1983asymptotic}
Iben~Jr I.,  Renzini A.,  1983, IN: Annual review of astronomy and astrophysics. Volume 21 (A84-10851 01-90). Palo Alto, CA, Annual Reviews, Inc., 1983, p. 271-342., 21, 271

\bibitem[\protect\citeauthoryear{Javadi, van Loon  \& Mirtorabi}{Javadi et~al.}{2011a}]{javadi2011auk}
Javadi A.,  van Loon J.~T.,   Mirtorabi M.~T.,  2011a, Monthly Notices of the Royal Astronomical Society, 411, 263

\bibitem[\protect\citeauthoryear{Javadi, van Loon  \& Mirtorabi}{Javadi et~al.}{2011b}]{javadi2011uk}
Javadi A.,  van Loon J.~T.,   Mirtorabi M.~T.,  2011b, Monthly Notices of the Royal Astronomical Society, 414, 3394

\bibitem[\protect\citeauthoryear{Javadi, van Loon, Khosroshahi  \& Mirtorabi}{Javadi et~al.}{2013}]{javadi2013uk}
Javadi A.,  van Loon J.~T.,  Khosroshahi H.,   Mirtorabi M.~T.,  2013, Monthly Notices of the Royal Astronomical Society, 432, 2824

\bibitem[\protect\citeauthoryear{Javadi, Saberi, van Loon, Khosroshahi, Golabatooni  \& Mirtorabi}{Javadi et~al.}{2015}]{javadi2015uk}
Javadi A.,  Saberi M.,  van Loon J.~T.,  Khosroshahi H.,  Golabatooni N.,   Mirtorabi M.~T.,  2015, Monthly Notices of the Royal Astronomical Society, 447, 3973

\bibitem[\protect\citeauthoryear{Javadi, Van~Loon  \& Khosroshahi}{Javadi et~al.}{2016}]{javadi2016role}
Javadi A.,  Van~Loon J.,   Khosroshahi H.,  2016, arXiv preprint arXiv:1610.00254

\bibitem[\protect\citeauthoryear{Javadi, van Loon, Khosroshahi, Tabatabaei, Hamedani~Golshan  \& Rashidi}{Javadi et~al.}{2017}]{javadi2017uk}
Javadi A.,  van Loon J.~T.,  Khosroshahi H.~G.,  Tabatabaei F.,  Hamedani~Golshan R.,   Rashidi M.,  2017, Monthly Notices of the Royal Astronomical Society, 464, 2103

\bibitem[\protect\citeauthoryear{Kacharov, Rejkuba  \& Cioni}{Kacharov et~al.}{2012}]{kacharov2012spectra}
Kacharov N.,  Rejkuba M.,   Cioni M.-R.,  2012, Astronomy \& Astrophysics, 537, A108

\bibitem[\protect\citeauthoryear{Khatamsaz, Abdollahi, Abdollahi, Javadi  \& Van~Loon}{Khatamsaz et~al.}{2024}]{khatamsaz2024investigating}
Khatamsaz F.,  Abdollahi M.,  Abdollahi H.,  Javadi A.,   Van~Loon J.,  2024, IAU General Assembly, p.~648

\bibitem[\protect\citeauthoryear{Kroupa}{Kroupa}{2001}]{kroupa2001variation}
Kroupa P.,  2001, Monthly Notices of the Royal Astronomical Society, 322, 231

\bibitem[\protect\citeauthoryear{Lee, Freedman  \& Madore}{Lee et~al.}{1993}]{lee1993tip}
Lee M.~G.,  Freedman W.~L.,   Madore B.~F.,  1993, Astrophysical Journal v. 417, p. 553, 417, 553

\bibitem[\protect\citeauthoryear{Marigo, Girardi, Bressan, Groenewegen, Silva  \& Granato}{Marigo et~al.}{2008}]{marigo2008evolution}
Marigo P.,  Girardi L.,  Bressan A.,  Groenewegen M.~A.,  Silva L.,   Granato G.~L.,  2008, Astronomy \& Astrophysics, 482, 883

\bibitem[\protect\citeauthoryear{Marigo et~al.,}{Marigo et~al.}{2017}]{marigo2017new}
Marigo P.,  et~al., 2017, The Astrophysical Journal, 835, 77

\bibitem[\protect\citeauthoryear{Mateo}{Mateo}{1998}]{mateo1998dwarf}
Mateo M.,  1998, Annual Review of Astronomy and Astrophysics, 36, 435

\bibitem[\protect\citeauthoryear{Navabi et~al.,}{Navabi et~al.}{2021}]{navabi2021isaac}
Navabi M.,  et~al., 2021, The Astrophysical Journal, 910, 127

\bibitem[\protect\citeauthoryear{Rezaei~kh, Javadi, Khosroshahi  \& van Loon}{Rezaei~kh et~al.}{2014}]{rezaei2014star}
Rezaei~kh S.,  Javadi A.,  Khosroshahi H.,   van Loon J.~T.,  2014, Monthly Notices of the Royal Astronomical Society, 445, 2214

\bibitem[\protect\citeauthoryear{Saremi, Javadi, Navabi, van Loon, Khosroshahi, Arbab  \& McDonald}{Saremi et~al.}{2021}]{saremi2021isaac}
Saremi E.,  Javadi A.,  Navabi M.,  van Loon J.~T.,  Khosroshahi H.~G.,  Arbab B.~B.,   McDonald I.,  2021, The Astrophysical Journal, 923, 164

\bibitem[\protect\citeauthoryear{Sibbons, Ryan, Cioni, Irwin  \& Napiwotzki}{Sibbons et~al.}{2012}]{sibbons2012agb}
Sibbons L.,  Ryan S.~G.,  Cioni M.-R.,  Irwin M.,   Napiwotzki R.,  2012, Astronomy \& Astrophysics, 540, A135

\bibitem[\protect\citeauthoryear{Sibbons, Ryan, Napiwotzki  \& Thompson}{Sibbons et~al.}{2015}]{sibbons2015spectral}
Sibbons L.,  Ryan S.~G.,  Napiwotzki R.,   Thompson G.,  2015, Astronomy \& Astrophysics, 574, A102

\bibitem[\protect\citeauthoryear{Tolstoy, Irwin, Cole, Pasquini, Gilmozzi  \& Gallagher}{Tolstoy et~al.}{2001}]{tolstoy2001using}
Tolstoy E.,  Irwin M.~J.,  Cole A.~A.,  Pasquini L.,  Gilmozzi R.,   Gallagher J.,  2001, Monthly Notices of the Royal Astronomical Society, 327, 918

\bibitem[\protect\citeauthoryear{Whitelock, Feast, Loon  \& Zijlstra}{Whitelock et~al.}{2003}]{whitelock2003obscured}
Whitelock P.~A.,  Feast M.~W.,  Loon J. T.~v.,   Zijlstra A.~A.,  2003, Monthly Notices of the Royal Astronomical Society, 342, 86

\bibitem[\protect\citeauthoryear{Whitelock, Menzies, Feast, Nsengiyumva  \& Matsunaga}{Whitelock et~al.}{2013}]{whitelock2013local}
Whitelock P.~A.,  Menzies J.~W.,  Feast M.~W.,  Nsengiyumva F.,   Matsunaga N.,  2013, Monthly Notices of the Royal Astronomical Society, 428, 2216

\bibitem[\protect\citeauthoryear{Zhang, Mackey  \& Da~Costa}{Zhang et~al.}{2021}]{zhang2021panoramic}
Zhang S.,  Mackey D.,   Da~Costa G.~S.,  2021, Monthly Notices of the Royal Astronomical Society, 508, 2098

\makeatother
\end{thebibliography}

\end{document}